\documentclass{aa}

\usepackage{natbib}
\usepackage{amsmath}
\bibpunct{(}{)}{;}{a}{}{,} 
\usepackage{graphics}   
\usepackage{graphicx}   
\usepackage{epstopdf}
\usepackage{longtable}   
\usepackage{url}      
\usepackage{bm}        
\usepackage[varg]{txfonts}
\usepackage{pdflscape}
\usepackage{supertabular}
\usepackage{hyperref}   
\usepackage[svgnames]{xcolor}
\usepackage{longtable}
\usepackage{lscape}
\usepackage{booktabs}
\usepackage{graphicx}
\usepackage{multicol}
\newcommand{\Msol}{M\ensuremath{_{\sun}}}

\usepackage{color}

\begin{document}

\title{Very massive star winds as sources of the short-lived radioactive isotope $^{26}$Al} 
\subtitle{}

\author{S\'ebastien Martinet\inst{1}, Georges Meynet\inst{1}, Devesh Nandal\inst{1}, Sylvia Ekstr{\"o}m\inst{1}, Cyril Georgy\inst{1}, Lionel Haemmerlé\inst{1}, Raphael Hirschi\inst{2,3}, Norhasliza Yusof\inst{4}, Matthieu Gounelle\inst{5}, Vikram Dwarkadas\inst{6}}

 \institute{ Geneva Observatory, University of Geneva, Chemin Pegasi 51, CH-1290 Versoix, Switzerland\\
 email: sebastien.martinet@unige.ch 
 \and Astrophysics Group, Keele University, Keele, Staffordshire, ST5 5BG, UK
 \and Institute for Physics and Mathematics of the Universe (WPI), University of Tokyo, 5-1-5 Kashiwanoha, Kashiwa 277-8583, Japan
 \and Department of Physics, Faculty of Science, University of Malaya, 50603 Kuala Lumpur, Malaysia
\and IMPMC, CNRS - UMR 7590, Sorbonne Université, Muséum national d’Histoire naturelle, 75005 Paris, France.
\and Department of Astronomy and Astrophysics, University of Chicago, 5640 S Ellis Ave, Chicago, IL 60637.}
 \authorrunning{Martinet et al.}

\date{Accepted May 2022}

\abstract{The $^{26}$Al short-lived radioactive nuclide is the source of the observed galactic diffuse $\gamma$-ray emission at 1.8 MeV. While different sources of $^{26}$Al have been explored, such as AGB stars, massive stars winds, and supernovae, the contribution of very massive stars has never been studied. }
{We study the stellar wind contribution of very massive stars, \textit{i.e} stars with initial masses between 150 and 300 M$_\odot$, to the enrichment in $^{26}$Al of the galactic interstellar medium.}
{We discuss the production of $^{26}$Al by studying rotating and non-rotating very massive stellar models with initial masses between 150 and 300 M$_\odot$ for metallicities Z=0.006, 0.014, and 0.020. We confront this result to a simple Milky Way model taking into account both the metallicity and the star formation rate gradients.}
{We obtain that very massive stars in the {{ Z=0.006-0.020}} metallicity range might be very significant contributors to the $^{26}$Al enrichment of the interstellar medium. Typically, the contribution of {the winds of massive stars} to the total quantity of $^{26}$Al in the Galaxy increases by 150\% when very massive stars are considered.
}
{Very massive stars,
despite their rarity, might be important contributors to $^{26}$Al and overall very important actors for nucleosynthesis in the Galaxy. 
}


\maketitle

\section{Introduction}
$^{26}$Al holds a special position among the short-lived radioisotopes with lifetimes shorter or equal to about 1 Myr. { It is, with $^{60}$Fe, the only element emitting a gamma-ray line that has been observed as a diffusive emission coming from the disk of our Galaxy. The origin of the gamma-ray emission is the decay of the} ground state of
$^{26}$Al with a half-life of 0.717 Myr { \citep{Norris1983}} into an excited state of $^{26}$Mg, the de-excitation of which emits a 1.805 MeV photon.

The emission line was first detected by the HEAO-C satellite \citep{Mahoney1982, Mahoney1984}, then confirmed by balloon born experiments \citep{Varen1992} and later-on by the ACE satellite \citep{Yanasak2001}. The first map of the Milky Way in this line has been obtained by the CGRO Comptel \citep{Ball1987, Chen1995}.
Subsequent new maps have been obtained by INTEGRAL \citep{Diehl2003}.
Studying emission line intensity and adopting reasonable assumptions about the distribution of $^{26}$Al in the Galaxy, a total amount between 1.7 and 3.5 M$_\odot$ of $^{26}$Al has been estimated to be present in the Milky Way today
\citep{KnoT1999, DiehlAA2006, Wang2009}, with current best estimate of 2 \Msol\ \citep{Pleintinger2020}.
Interestingly, a comparison between the 1.8 MeV intensity map and other maps at different wavelengths showed that the best correspondence is obtained with the free-free emission \citep{Knoff1999}. This clearly points to ionized regions where hot stars are present, and thus favors short-lived massive stars as the main contributors to this $^{26}$Al.


The { amount} of galactic $^{26}$Al likely remains constant with time, since we can reasonably assume that over the whole Galaxy and on a timescale that { covers} a few $^{26}$Al lifetimes, \textit{i.e.} a few Myr, a stationary equilibrium is reached between the production and destruction rate of this isotope. { Indeed, the change of the mass of $^{26}$Al in the Galaxy, $M_{26}$, can be written
$$
{{dM_{26} \over dt}=P_{26}-{M_{26}\over \tau_{26}}},
$$
where $P_{26}$ is the rate of production of $^{26}$Al and $M_{26}\over \tau_{26}$ its decay rate with $\tau_{26}$ the decay { constant}. We see that when, for instance, $M_{26}$ is so low that the right-hand side is positive, then $^{26}$Al will accumulate in the ISM. In the reverse case, $M_{26}$ is so high that the right-hand side is negative. Thus, on timescales of a few $\tau_{26}$, an equilibrium is reached between the production and the destruction rate of $^{26}$Al. This implies that at any time, one can reasonably assume that
$M_{26} \sim P_{26}\tau_{26}$ \citep[see the question about the granularity of nucleosynthesis in][]{Meyer2000}.
The $^{26}$Al
whose decay can be observed as a diffuse $\gamma$-ray emission in the Galaxy today needs to have been ejected into the interstellar medium in the last million years.}
Thus, the 1.8 MeV observation is a measure of the recent nucleosynthetic activity (of at least some specific sources) in our Galaxy \citep[see e.g. the reviews by][]{Diehl2021ISM}. 


{ To estimate $P_{26}$, we must identify the sources of $^{26}$Al.}
Massive stars can contribute { to the $^{26}$Al budget} through their winds and at the moment of their explosion as a supernova. Grids of models predicting the quantities ejected by stellar winds \citep{PC1986, Meynet1997,Palacios2005}, by winds and supernovae \citep{LC2006}, and more recently by winds in close binaries \citep{Brinkman2019} have been published. 
Other sources have been explored such as supermassive stars \citep[stars above 10$^4$\Msol][]{Hillebrandt1987}, Asymptotic Giant Branch (AGB) stars \citep{Wasserburg2006,NM2005} and novae \citep{Bennett2013}. The { AGB}, however, are considered nowadays as likely minor contributors.

Based on the prevalence of massive young stars as $^{26}$Al sources, models for specific young star forming regions like the Carina and the Cygnus regions for which maps can be obtained \citep{Kno1996,Kno2002, Bou2003}, or more generally for starbursts regions  have also been built \citep{delRio1996,Cervin2000,Higdon2004,Roth2006, Voss2009, Lacki2014, Krause2015}, showing a reasonable agreement with observations.
Of course, a still better constraint would be to see the emission of one source. Some trials have been done to detect a signal from the nearest Wolf-Rayet star $\gamma$ Velorum \citep{Oberlack2000} but without success. In case of a point-like source, models would however predict enough flux to be detectable. An explanation for the non-detection might be that due to the velocities of the stellar winds, {the $^{26}$Al is actually distributed rapidly enough in a large area around the star, which would then make the detection much harder \citep{NM2005, NMGM2006}. Indeed, the minimum level for detecting an emission is larger for a source diluted over a large area than for a point-like source.}

Recent papers, that have simulated galactic-scale $^{26}$Al maps \citep{Plein2019}, have also explored the contribution of massive stars towards $^{26}$Al and $^{60}$Fe at the scale of the Galaxy \citep{Wang2020}. Furthermore, questions such as how $^{26}$Al can trace metal losses through hot chimneys \citep{Krause2021},
or how it is distributed in superbubbles \citep{Rodgers2019} have also been explored. All these works are based on $^{26}$Al masses ejected by stars with initial masses between 25 and around 120 M$_\odot$. 
{ However, stars more massive than 120 M$_\odot$ up to at least $\sim$ 300\Msol\ likely exist \citep{Crowther2010,Bestenlehner2020,Brands2022} and their extreme evolution and large mass loss rates could make them important contributor to the $^{26}$Al production in the Galaxy.}
In the present paper, we explore the contribution of these very massive stars { in enriching the interstellar medium in $^{26}$Al through their winds}. Although these stars are very rare (see Sect. \ref{sect:conlusion_perspectives}), depending on the mass loss rates they exhibit, their contribution might be important. 


 The paper is organized as follows: in Section 2, we briefly indicate the physical ingredients used to compute the present models for very massive stars. The physics contributing to very massive stars being $^{26}$Al potential sources is discussed in Section 3. 
The $^{26}$Al masses ejected by the stellar winds of these stars and their dependence on initial mass, rotation. and metallicity is discussed in Section 4. A simple estimate of the very massive stars contribution to the total $^{26}$Al mass budget in the Milky Way is given in Section 5. The main conclusions are synthesized in Section 6 with some possible links with other topics involving $^{26}$Al.

\section{Ingredients of the stellar models}

{ We computed very massive stars models with GENEC \citep{Eggenberger2008} with initial masses of 180, 250 and 300\Msol\ for Z=0.006 and Z=0.014, without rotation and with a rotation rate of V/V$_c$=0.4 where V$_c$ is the critical velocity\footnote{ The critical velocity is the velocity such that the centrifugal force at the equator balances the gravity there. Its expression is taken as indicated by expression 6 in \citet{Eks2008}.}. The nuclear network allows following the variation of abundances of 30 isotopes \footnote{These isotopes are  $^{1}$H, $^{3,4}$He, $^{12,13,14}$C, $^{14,15}$N, $^{16,17,18}$O, $^{18,19}$F, $^{20,21,22}$Ne, $^{23}$Na, $^{24,25,26}$Mg, $^{26,27}$Al, $^{28}$Si, $^{32}$S, $^{36}$Ar, $^{40}$Ca, $^{44}$Ti, $^{48}$Cr, $^{52}$Fe and $^{56}$Ni.}. In addition to the CNO cycles, the Ne-Na and Mg-Al chains are included. Note that the 
the isomeric ($^{26}$Al$^{m}$) and the ground state ($^{26}$Al$^{g}$) of $^{26}$Al are considered as two different species.} The nuclear reaction rate of $^{25}$Mg(p,$\gamma$)$^{26}$Al$^{g,m}$ is taken from \citet{Iliadis2001}.  \citet{Champagne1993} is used for $^{26}$Al(p,$\gamma$) $^{27}$Si and \citet{CF88} for $^{26}$Al(n,$\alpha$) $^{23}$Na and $^{26}$Al(n,p)$^{26}$Mg. 
Note that the rates are the same as those used in the previous study on that topic by \citet{Palacios2005}.

The present models differ mainly in three points (Martinet et al. in prep) with respect to the physics used in the grids by \cite{Ekstrom2012} and \cite{Yusof2022}. First, we adopted the Ledoux criterion for convection
instead of Schwarzschild and an overshoot of 0.2 $H_p$ instead of 0.1 $H_p$. These changes were done since there are some indications that the Ledoux criterion might be more appropriate \citep[][]{Georgy2014, Kaiser2020} { for these stars, and that an increase of the overshoot parameter is needed for stars larger than 8\Msol\ \citep{Martinet2021}}. Another change is that the present very massive star models have been computed with an equation of state accounting for electron-positron pair production \citep{Timmes2000}\footnote{The models at $Z$=0.020 have been computed with the same equation of state as in \cite{Ekstrom2012}}. These changes have, however, very little impact on the question discussed in the present paper. Indeed, the results depend on the structure of the models during the Main-Sequence (MS) phase. Changing from Schwarzschild to Ledoux during the MS phase does not change anything, since the mass of the convective core decreases with time, producing no gradient of chemical composition in the layer just above the convective core. A larger overshoot will tend to reduce the time between the beginning of the evolution and the first surface enrichment, in $^{26}$Al. However, very massive stars have anyway very large convective cores and this time would be short even with a smaller overshoot. The change of the equation of state does not have any effect on the MS phase.


The radiative mass-loss rate adopted on the MS is from \citet{Vink2001}; the domains not covered by this
prescription { \citep[see Fig. 1 of][]{Eggenberger2021}} use the \citet{deJager1988} rates. 
{ \citet{Grafener2008} prescriptions are used in their domain of application, while \citet{Nugis2000} prescriptions are used everywhere else} for the Wolf-Rayet phase. The Wolf-Rayet phase is assumed to begin 
when the model has an effective temperature larger than 10 000 K and a surface mass fraction of hydrogen at the surface below 0.3.
The radiative mass-loss
rate correction factor described in \citet{Maeder2000} is applied for rotating models. 
{ The dependence on metallicity is taken such that  $\dot{M}(Z) = (Z/Z_\odot)^{0.7} \dot{M} (Z)$, except during
the red supergiant (RSG) phase, for which no dependence on the metallicity is used. This is done accordingly to \citet{vanLoon2005} and \citet{Groenewegen2012a,Groenewegen2012b} showing that the metallicity dependence for the mass loss rates of these stars do appear to be weak.}
For the rotation prescription, the models use the shear diffusion coefficient as given by \cite{Maeder1997} and the horizontal diffusion coefficient from \cite{Zahn1992}.



\section{$^{26}$Al production and wind ejection in very massive stars}

\begin{figure*}[!h]
    \centering
    \includegraphics[width=0.49\textwidth]{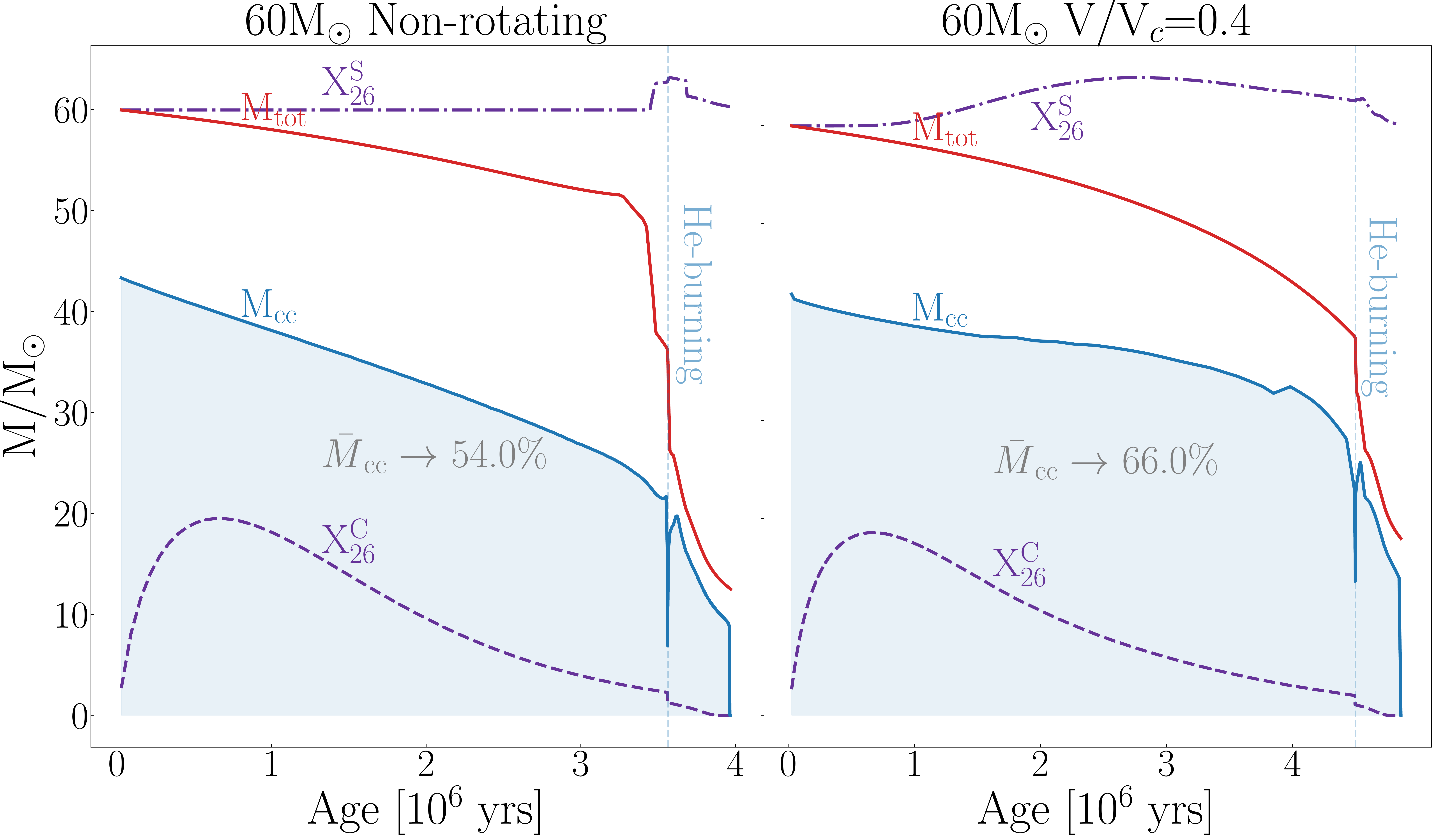}
    \includegraphics[width=0.49\textwidth]{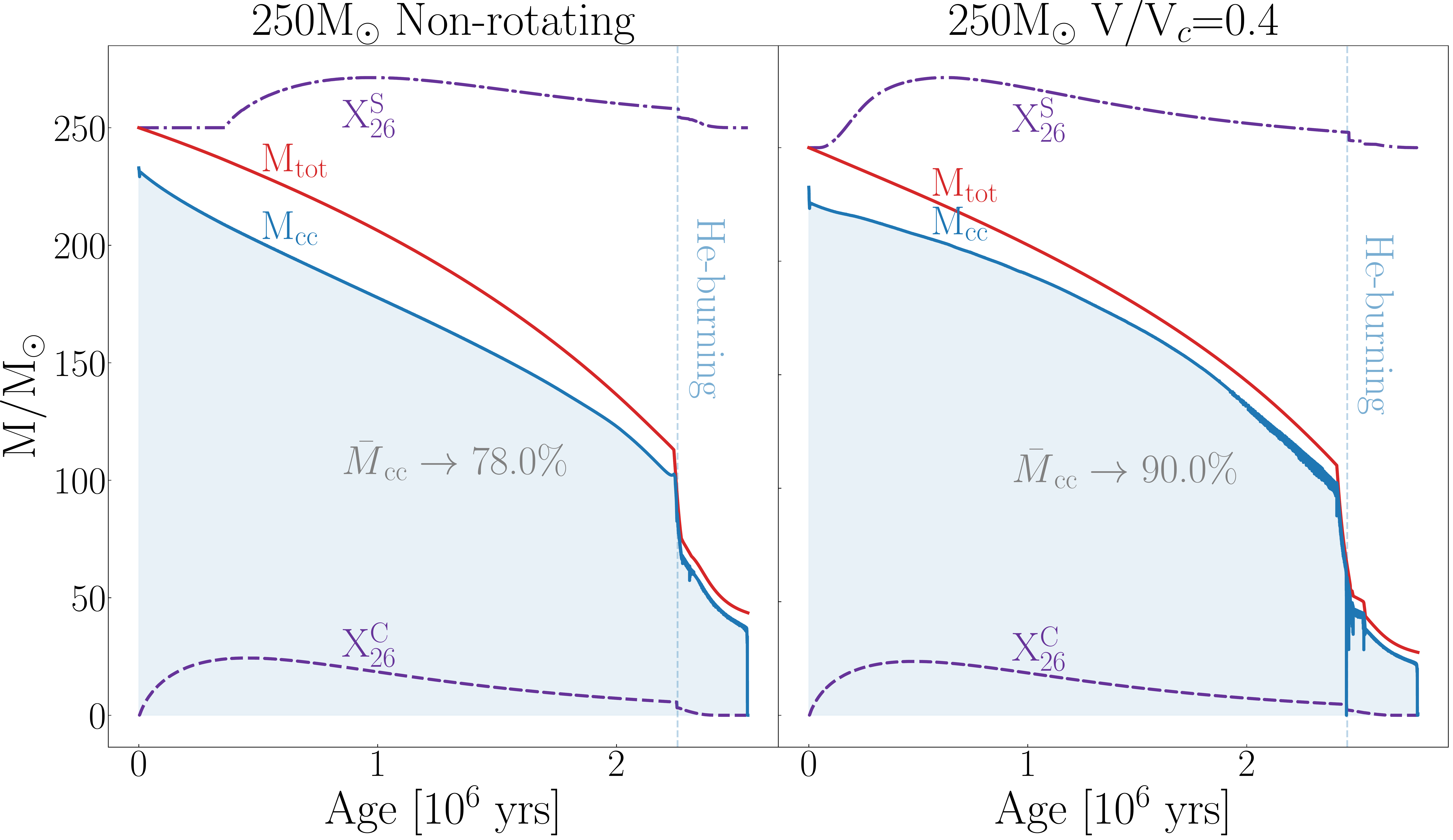} 
     \caption{Evolution as a function of the stellar age in Myr of the total mass (M$_{\rm tot}$, see the continuous red curve), of the convective core mass (M$_{\rm cc}$, continuous blue curve) in solar mass units, of the central (X$_{26}^{\rm C}$, dashed purple line) and surface $^{26}$Al abundance (X$_{26}^{\rm S}$, dashed-dotted purple line) in mass fractions for a non-rotating and a rotating stellar model at a metallicity Z=0.014. The initial rotation of the rotating models is 40\% the critical velocity at the ZAMS. For clarity purpose, the curves showing the $^{26}$Al mass fraction are multiplied by factors equal to 7.0$\times$10$^5$ and 10$^6$ for respectively the central and surface values. The curve for the surface value is shifted so that its starting point at an age equal to 0 is 60 and 250 in the left and right panels respectively. { The time-averaged convective core mass fraction during the MS phase is indicated. The light blue vertical line indicates the end of the core H-burning phase and the beginning of the core He-burning one.} }    
  \label{fig:kipp}

\end{figure*}

The evolution as a function of the age for 60 and 250 M$_\odot$ models are shown in the left and right panel of Fig.~\ref{fig:kipp} respectively. As is well known, $^{26}$Al is synthesized during the core-H burning phase by proton capture on $^{25}$Mg. We see that in the 60 M$_\odot$, the abundance of $^{26}$Al at the center reaches a maximum before the age of 1 Myr. The decrease that follows results from the fact that the $\beta$-decay of the radioisotope dominates over the production process when the abundance of $^{25}$Mg decreases. We see that at the beginning of the core He-burning phase, there is a rapid drop in the central $^{26}$Al mass fraction. $^{26}$Al is destroyed at the beginning of the core He-burning phase mainly by neutron captures, { initially released by the $^{13}$C$(\alpha,n)^{16}$O reactions, and then mainly by { $^{22}$Ne($\alpha$,n)$^{25}$Mg}}. 













During the core H-burning phase,
as a result either of mass loss alone (in non-rotating models) or of mass loss and rotationally induced mixing in the outer radiative zone, the surface, and hence the winds, become enriched in $^{26}$Al. This surface enrichment lasts until products of core He-burning appear at the surface. From this stage on, the surface abundance of $^{26}$Al rapidly decreases,{ reflecting the destruction of $^{26}$Al when interior regions of the star processed by He-burning reactions are exposed at the surface by stellar winds.}

Comparing the rotating and the non-rotating model for the 60 M$_\odot$, we see, as was already discussed in \citet{Palacios2005} that rotation favors the $^{26}$Al wind enrichment through the following effects{ :}
\begin{itemize}
    \item Species synthesized in the core { appear at the surface by rotational mixing in a timescale that is shorter than the time for mass loss to uncover layers whose composition has been changed by nuclear burning}. Typically, looking at the left panel of Fig.~\ref{fig:kipp}, we see that the maximum abundance at the surface is reached at an age of $\sim$3.6 Myr in the non-rotating 60 M$_\odot$ model, while the same surface mass fraction is reached at an age of less than 2 Myr in the corresponding rotating model.
    \item Due to diffusion of hydrogen from the radiative envelope into the convective core, the convective core remains larger in the rotating model than in the non-rotating one. This also favors a rapid emergence of core H-burning products at the surface.
    \item The core H-burning lifetime is increased by rotation from 3.5 to more than 4.5 Myr in the case of the 60 M$_\odot$ model, which supports $^{26}$Al wind ejection during that phase.

\end{itemize}
 We can wonder whether the diffusion of some $^{25}$Mg from the radiative envelope into the convective core contributes in some significant way to the increase of $^{26}$Al produced in the rotating model. We note that the maximum value of the mass fraction of $^{26}$Al at the center of the rotating 60 M$_\odot$ is slightly above the maximum value reached in the non-rotating corresponding model. This may be in part due to that effect, but also due to the difference in the central temperatures between the two models (see the discussion in Sect.~4).
 
Looking now at the right panel of Fig.~\ref{fig:kipp} that shows the case of a 250 M$_\odot$ very massive star, we can note the following differences with respect to a more classical 60 M$_\odot$ stellar model:
\begin{itemize}
\item $^{26}$Al appears at the surface significantly earlier, typically at ages of a few 0.1Myr. This comes mainly from the fact that convective cores in very massive stars occupies a much larger fraction of the total mass than in less massive stars. 
\item The time difference between the reaching of the peak abundance at the center and at the surface is reduced, thereby further reducing the time for the decay of $^{26}$Al between these two epochs. This favors larger ejected amounts.
\item The mass lost with a composition bearing the signatures of H-burning corresponds to 60\% of the initial mass for the 250 M$_\odot$ model, while it corresponds to 25\% in the case of the rotating 60 M$_\odot$ and less for the non-rotating one. This reflects how the mass loss increases with the initial mass.
In the present models we likely underestimate the mass loss rates and thus the present predictions are likely on the conservative side (see the discussion in Sect.~6). 
\item The impact of rotation, while non-negligible, is not as strong in the 250 M$_\odot$ stellar model as in the 60 M$_\odot$ one. This reflects the fact that mass loss by stellar winds and convection dominates in a more important manner the evolution of those stars compared to lower initial mass models.
\end{itemize}

{ Overall, rotation has a major effect on the increase of the $^{26}$Al yields, as strong transport from rotational mixing brings large quantities of $^{26}$Al to the surface earlier-on, enabling the $^{26}$Al enriched envelope to be lost by the winds on longer timescales. Larger initial masses, while of course containing a larger reservoir of $^{25}$Mg to produce $^{26}$Al, will also have stronger mixing due to larger convective cores. This effect, combined with the larger mass loss events, will be predominant over the rotation effect when scaling to very high initial masses. }
\section{The $^{26}$Al stellar yields}

For each model, we can compute the quantity of $^{26}$Al that is ejected by stellar winds, $Y_{\rm Al26}^{\rm winds}$. This quantity is obtained computing the integral below
\begin{align}
    Y_{\rm Al26}^{\rm winds}(M,Z,V)=\int_{0}^{\tau (M, Z, V)} X_{26}^{\rm S}(M, Z, V,t) \dot{M}(M,Z,V,t) dt
    \label{Eq:stellar_yields}
\end{align}
where $\tau (M, Z, V)$ is the lifetime of a star with an initial mass $M$, an initial metallicity $Z$ and an initial rotation $V$, $X_{26}^{\rm S}$ and $\dot{M}$ are the mass fraction of $^{26}$Al at the surface and the mass loss rates for the same model as a function of time.
{ While the decay of $^{26}$Al in the stellar interior is of course accounted for}, we are not accounting for the $^{26}$Al decay in the wind ejecta, because as explained in the { introduction,} we need to evaluate the production rate of $^{26}$Al. 

\begin{figure*}
\centering
     \includegraphics[width=0.75\textwidth]{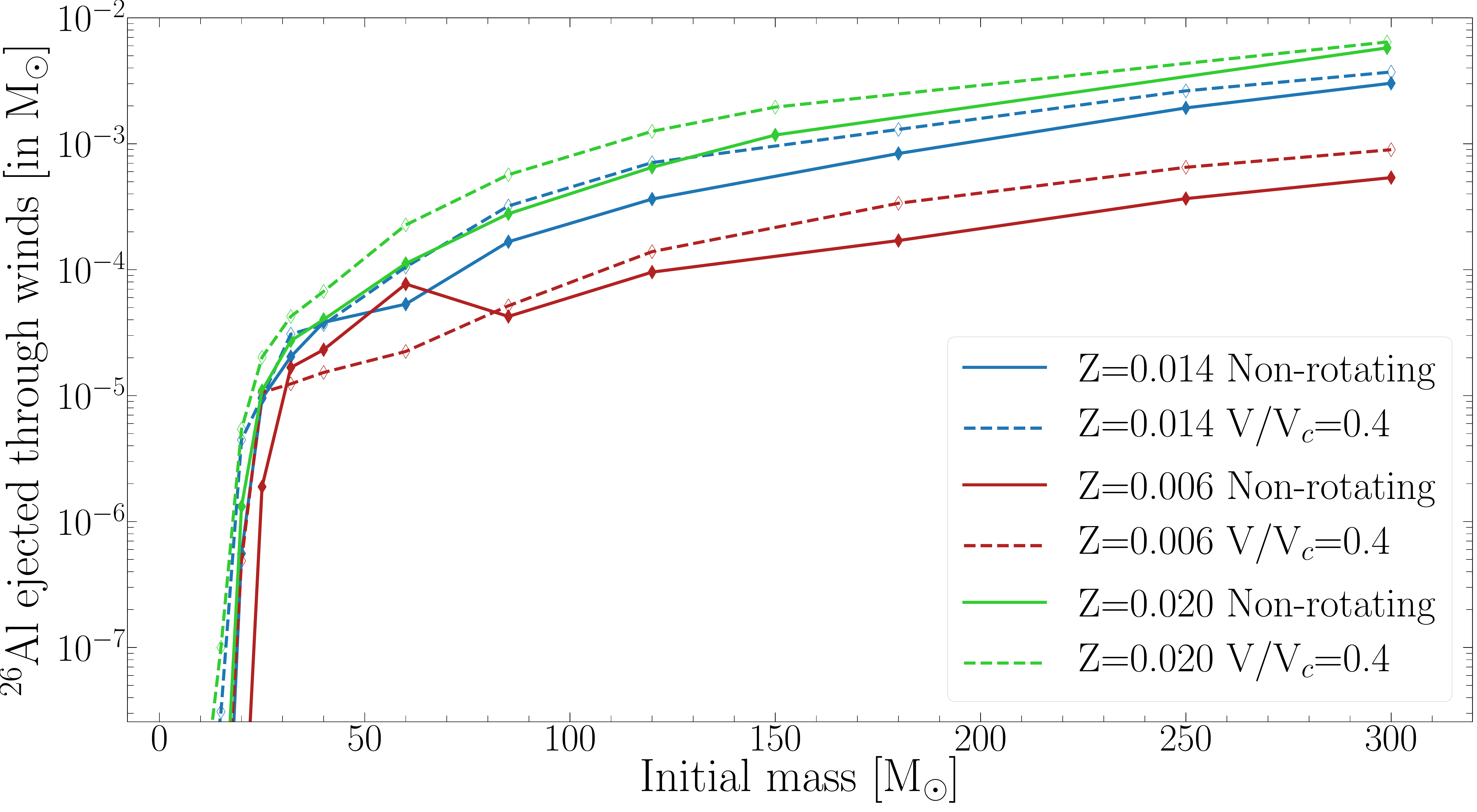}
     \caption{Masses of $^{26}$Al ejected by stellar winds during the total stellar lifetime of massive stars with different initial masses, initial metallicities and rotation rates. 
     }
     \label{fig:yields}
\end{figure*}

The stellar yields resulting from our different models { can be found in Table \ref{tab:Al26_production} in the appendix and} are shown as a function of the initial mass, initial metallicity and rotation in Fig.~\ref{fig:yields}. We see that the { yields} become larger than 10$^{-5}$ M$_\odot$ for initial masses above about 25 M$_\odot$. In general, above this mass, the $^{26}$Al increases with the mass, metallicity, and rotation. We can note the following interesting features in Fig.~\ref{fig:yields}:
\begin{itemize}
    \item For initial masses equal to or above 60 M$_\odot$ $Y_{\rm Al26}^{\rm winds}$ increases with the { initial} mass and the metallicity according to power laws. Typically, for solar metallicity models with rotation, $Y_{\rm Al26}^{\rm winds} \propto M^{1.93}$, and for a rotating 120 M$_\odot$, $Y_{\rm Al26}^{\rm winds} \propto Z^{1.45}$.
    \item Above 60 M$_\odot$, the differences between rotating and non-rotating models decreases, reflecting the fact that, in the very high mass regime, models are dominated by mass loss and convection more than by rotation (at least for the rotational velocities considered here).
    \item Below 60 M$_\odot$, we note that the behavior { of the yields} as a function of mass can be non-monotonous (see the non-rotating $Z$=0.006 models). Typically, at this metallicity, the yield of the 60 M$_\odot$ model presents a local maximum. This is due to the specific combination of the
    effects of stellar winds and convection in this model.  { This combination at 60 M$_\odot$}  produces
    a longer phase { than in higher initial mass models, during which layers processed by H-burning appear at the surface}. 
    \item The non-rotating models at $Z$=0.020 give similar yields as the rotating ones at $Z$=0.014
    showing thus a degeneracy of the yields between increase in rotation and increase in metallicity.
    \item The rotating models at $Z$=0.006 give lower yields than the non-rotating models for masses below about 85 M$_\odot$.
    This is due to the fact that non-rotating Z=0.006 models from 40 to 85 \Msol\ remain at lower effective temperature than rotating models during He-burning. This induces larger mass losses for these models and  results in larger $^{26}$Al yields.
\end{itemize}

\begin{figure}
     \includegraphics[width=0.49\textwidth]{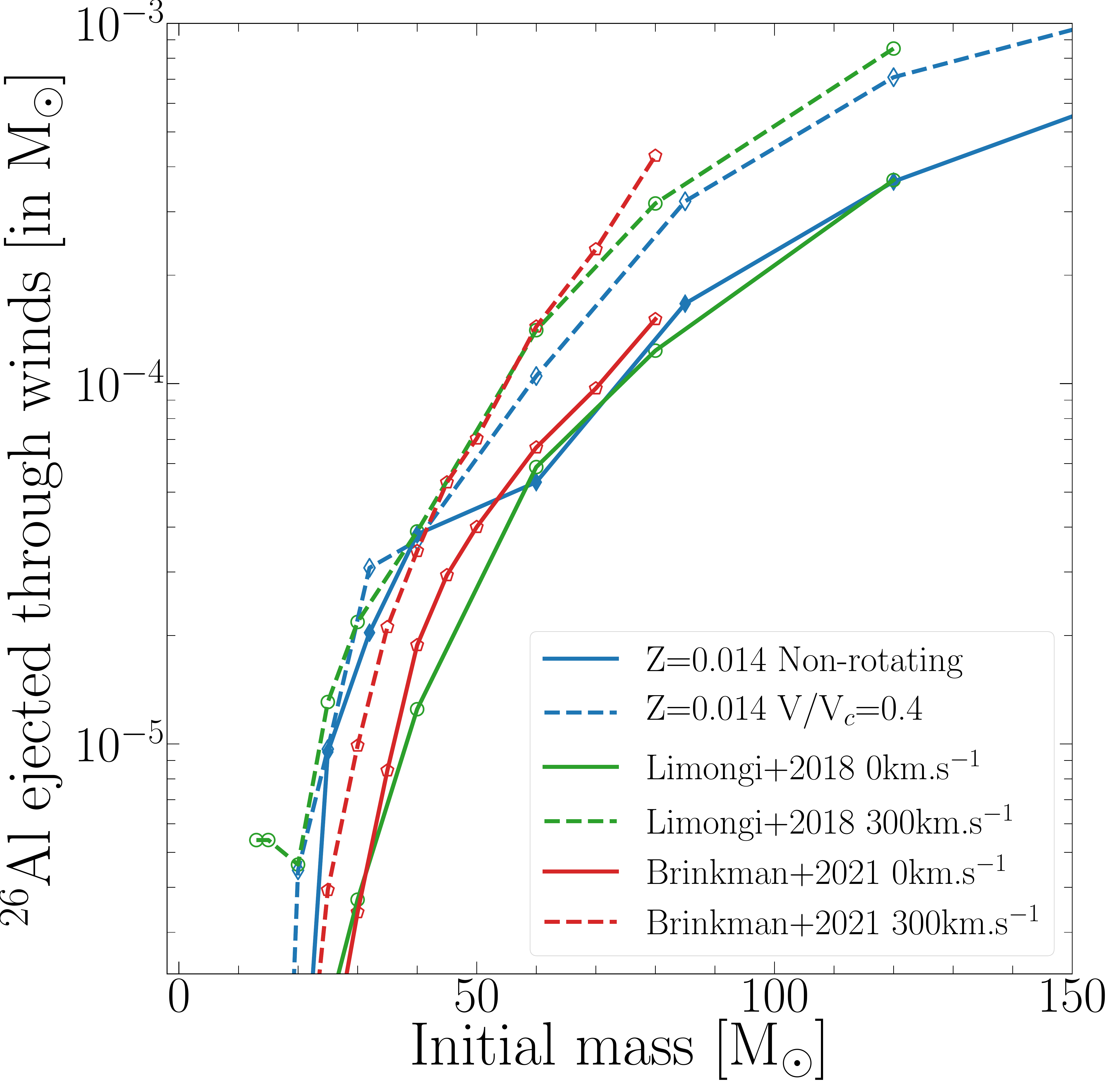}
     \caption{Plot of yields of present models at Z=0.014 in comparison to \citet{Limongi2018} and \citet{Brinkman2021} single star models. At 50M\Msol, V/V$_c$=0.4 is equivalent to 300 km s$^{-1}$, and at 150\Msol\ to 400 km s$^{-1}$.}
     \label{fig:yields2}
\end{figure}

Figure~\ref{fig:yields2} compares the stellar yields obtained from the present models with and without rotation at $Z$=0.014
with models by different authors. { Both \citet{Limongi2018} and \citet{Brinkman2021} models displayed here are for single solar metallicity stars.} 
For initial masses above 60 M$_\odot$, there is a general good agreement between the predictions of the different models for both rotating and non-rotating models.
In the 30 to 60 \Msol\ mass range, the non-rotating models of this work produce more $^{26}$Al than \citet{Limongi2018} and \citet{Brinkman2021} models, while the rotating models agree. 
{ Below 25 M$_\odot$, the present yields are larger than both \citet{Brinkman2021} and \citet{Limongi2018} yields due to larger mass loss in RSG stars, except for the 12 to 15 \Msol\ mass range, where a large difference is displayed by the \citet{Limongi2018} models.} This is due to the higher mass loss rates chosen for these models. They have implemented rates obtained from dust-driven wind during the RSG phase \citep[see][]{Chieffi2013}. 
Except for this mass domain, models present an overall good agreement, with differences that remain at a moderate level, or at least at a level that cannot be discriminated by any observations at the moment.


\section{Contribution of the winds of very massive stars to $^{26}$Al in the Milky Way}

\begin{figure*}[h]
    \centering
     \includegraphics[width=0.85\textwidth]{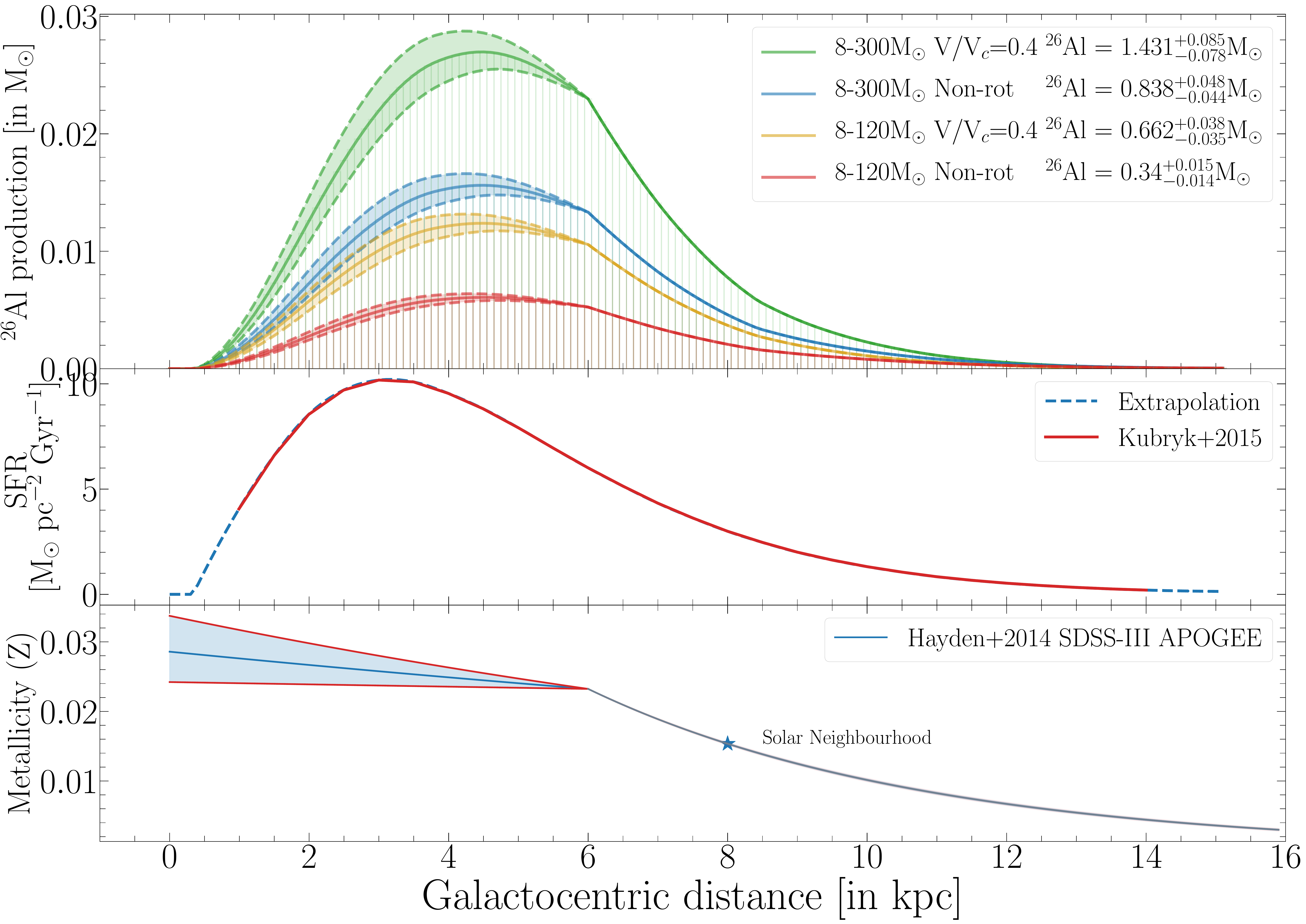}
     \caption{\textit{Top panel: }$^{26}$Al production in rings of 0.1 kpc width as a function of the galactocentric distance. The production is displayed for a Salpeter's IMF ($\alpha$=2.35) with different maximum mass and with models with and without rotation. The results are obtained taking into account the SFR and the metallicity radial gradient in the Milky Way. The dashed lines show the result when considering the uncertainties on the metallicity displayed in the bottom panel. The integrated quantities over the whole Milky Way of $^{26}$Al are given in the legend. \textit{Middle panel: }Star formation rate as a function of the galactocentric distance from \citet{Kubryk2015}. \textit{Bottom panel: }Metallicity as a function of the galactocentric distance from \citet{Hayden2014}.}
     \label{fig:Al26_galaxy}
\end{figure*}

In this { section,} we compute the quantity of $^{26}$Al present in the Milky Way due to massive star stellar winds. 
Let us call $\Psi (R)$ the Star Formation Rate (SFR) given in terms of the number of stars forming per unit surface, and per unit time at a given galactocentric radius, $R$, in the Galaxy. We assume that this function is constant in the last 10 Myr and that the Galaxy is axisymmetric. Note that our objective here is to estimate the impact of including very massive stars in the $^{26}$Al due to stellar winds, rather than to provide a very detailed estimate based on a more complex model for the Galaxy.
{ We normalize the initial mass function, $\Phi(M)dM$, such that 
\begin{align}
    \int_{0.07M_\odot}^{300M_\odot}\Phi(M)dM=1.0.
\label{Eq:Normalization}
\end{align}
\noindent Thus we can interpret $\Phi(M)dM$  as the probability that a star of initial mass between $M$ and $M+dM$ is formed when stars with initial masses between 0.07 and 300 M$_\odot$ are formed.
} We assume that this function does not vary with { time}, the SFR, neither with the metallicity, $Z$.
Let us estimate the production rate of $^{26}$Al by stellar winds of stars of a given age $t$, with an initial mass between $M$ and $M+dM$, a initial metallicity $Z$ and initial rotation $V$ at a given galactocentric distance $R$, per unit surface. This quantity is given by $\Psi(R)\times\Phi(M) dM\times X_{26}^{S}(M,Z,V,t)\dot{M}(M,Z,V,t)$.
The contribution of such stars irrespective of their age is obtained by integrating this expression over a period
that begins at a time $t_0-\tau(M,Z,V)$ and $t_0$ thus over a period whose duration is equal to the total lifetime of the star considered. { The contribution of stars of a given mass $M$, whatever their age,} is given by
$\Psi(R)\times\Phi(M)\times Y_{\rm Al26}^{\rm winds}$. The contribution of stars of different masses is {then} given by $\Psi(R) \bar Y_{\rm Al26}^{\rm winds}$, where 
\begin{align}
    \bar Y_{\rm Al26}^{\rm winds}=\int_{M_{\rm Min}}^{M_{\rm Max}} Y_{\rm Al26}^{\rm winds}(M,Z,V) \Phi(M) dM
\end{align}
is the { average} mass of $^{26}$Al ejected by stellar winds per star formed { in a given initial mass interval}.
The rate of $^{26}$Al production in a galactic ring with an internal radius $R$ and width d$R$ is given by
$2 \pi R \Psi(R) \bar Y_{\rm Al26}^{\rm winds}$d$R$ and the total mass in the Galaxy is obtained by
integrating the above expression over the whole plane of the Milky Way. Let us call this quantity P$_{26}$, the galactic production rate. Let us note that the different rings have different metallicities, since the Galaxy presents a metallicity gradient such that the metallicity tends to increase when the galactocentric distance decreases. The above integration needs to account { for this, such that the yields corresponding to the metallicity of each ring depending on its distance to the galactic center are adopted.} From this { production rate}, we can estimate the total amount of $^{26}$Al in the Galaxy due to stellar winds, M$_{26}$, assuming that this total amount does not depend on time. As seen in the introduction, it is equal to M$_{26}$=P$_{26}\tau_{26}$.


In Fig.~\ref{fig:Al26_galaxy}, we show the result of such an integration adopting the star formation rate from \citet{Kubryk2015}\footnote{To obtain from this star formation rate, given in solar mass per pc$^{-2}$ and per Gyrs (see middle panel of Fig.~\ref{fig:Al26_galaxy}), a star formation rate given in number of stars per pc$^{-2}$ and per Gyrs, we need to divide by the average mass of a star when stars are formed over the whole mass range between 0.07 and 300 M$_\odot$.}, slightly reduced to obtain a number of core-collapse supernovae of 2 per century in the Milky Way, a Salpeter's IMF with $\alpha$=2.35 and a metallicity gradient taken from \citet{Hayden2014} (shown in the bottom panel of
Fig.~\ref{fig:Al26_galaxy}).

The top panel of Fig.~\ref{fig:Al26_galaxy} shows the $^{26}$Al production in rings of 0.1 kpc width as a function of the galactocentric distance. { We consider four different sets of stellar populations, with and without rotation, and with and without the inclusion of very massive stars (VMS) in the IMF. These four sets have been chosen to underline the effect of rotational mixing on the galactic production of $^{26}$Al, and to explore the impact on $^{26}$Al production when including only a few VMS in these populations. The choice of a maximum mass of 300\Msol\ is motivated by the largest initial masses of VMS derived from observations of the Tarantula nebula   \citep{Bestenlehner2020,Brands2022} in the Large Magelanic Cloud (LMC). The resulting total amount of $^{26}$Al produced by such populations ranges from 0.340\Msol\ to 1.431\Msol.} This integrated quantities are displayed in the legend and give the total content of $^{26}$Al in the Milky Way produced by the winds according to the stellar models. { Only stars above 8\Msol\ are used to compute the $^{26}$Al lost by the winds. Indeed, lower mass stars have very limited mass loss and stars below 8-12\Msol\ do not even produce $^{26}$Al through the Ne-Na and Mg-Al chains due to their lower central temperature during core H-burning\footnote{ Note that AGB stars can reach high temperature in their H-burning shell and thus can contribute to the production of this element. However, this is another channel of production that is not discussed here.}. Of course, the number of stars above 8\Msol\ is computed through the SFR taking into account every star in the Milky Way model, hence normalized on the whole range from 0.07\Msol\ to the maximum mass included (here either 120\Msol\ or 300\Msol) as we have seen in Eq. \ref{Eq:Normalization}. } The extrapolation in metallicity has been made for each initial mass and are compatible with super solar mass loss rates. 

We can see that for every population models, the $^{26}$Al peak production is around 4.5kpc. It is due to the combination between three factors: (1) the peak of SFR, leading to larger quantities of massive stars, producing more $^{26}$Al; (2) the high metallicity of the inner part of the Milky Way, leading to higher yields as we have seen in Fig. \ref{fig:yields} and (3) the surface covered by each bin, indeed while the { bins are of 0.1kpc in width,} the surface they cover increases with $\pi$R$^2$. The combination of these three components explains the slight shift between the peak of SFR and the $^{26}$Al production. The $^{26}$Al production then diminishes more abruptly from 6kpc to the outer parts. This is directly linked to the change of trend in metallicity, dominating the yields. 

The impact of rotation can be seen when comparing the red (non-rotating) and the yellow (V/V$_c$=0.4) curves. As expected from Section 4, the higher yields produced by rotating models lead to a two times larger Galactic $^{26}$Al production. The larger effect is seen around the peak of SFR, once again due to the larger number of massive stars, for which all rotating models produce more $^{26}$Al at { these super solar metallicities}. 

The impact of including very massive stars can be seen when comparing the red (IMF going from 8 to 120\Msol) and the blue (8 to 300\Msol) curves. 
The inclusion of the VMS into the stellar population leads to a 120-150\% increase of the $^{26}$Al galaxy production. The larger effect is seen once again around the peak of SFR, where VMS have a higher probability to be produced. This means that { a few VMS are} sufficient to increase the $^{26}$Al production significantly. 

Finally, the combined effect of rotation and including the VMS leads to a four times larger $^{26}$Al production compared to models neither accounting for rotation nor VMS.
These results show that despite the fact that very massive stars are very rare (see Sect. \ref{sect:conlusion_perspectives} and Fig. \ref{fig:Number_VMS_galaxy}), their effect is still significant due to their large yields. Only a few VMS can have an important impact on the $^{26}$Al production at the Galactic scale.
This underlines the need to improve our knowledge about their frequency at various metallicities.

\begin{figure*}
    \centering
     \includegraphics[width=0.80\textwidth]{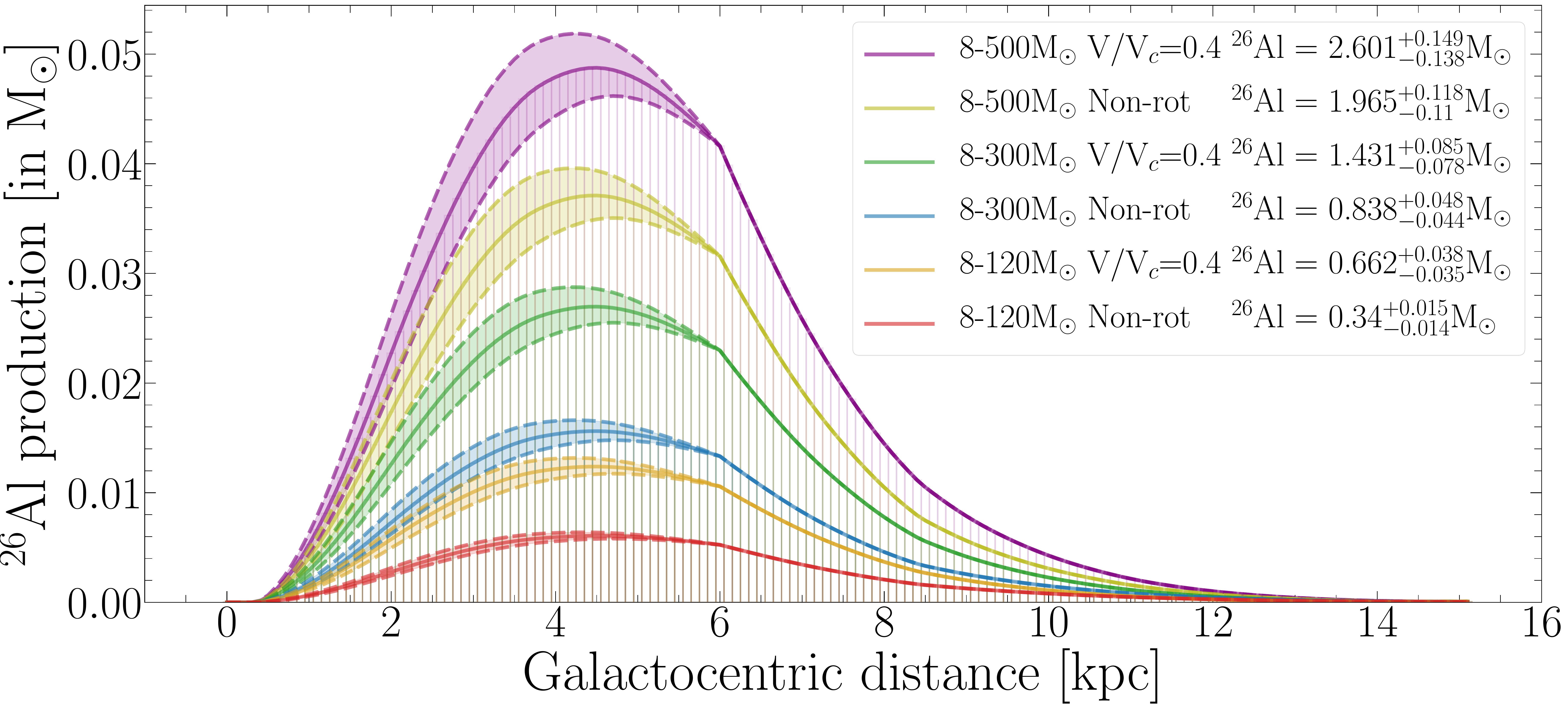}
     \caption{Effect of the upper-mass limit on the $^{26}$Al production in rings of 0.1 kpc width as a function of the galactocentric distance. The same methodology is used as in Fig. \ref{fig:Al26_galaxy}. The $^{26}$Al yields above 300\Msol\ are extrapolated.}
     \label{fig:upper_mass_limit}
\end{figure*}

\section{Discussion and conclusions}

\subsection{Impact of changing physical ingredients of the models}

{ The evolution of very massive star models depends} mainly on the mass loss rates.
In the present work, we likely underestimate the mass loss rates due to uncertainties on Eddington mass loss rates \citep{Vink2011, Besten2014,Vink2018} and hence the $^{26}$Al yields might also be
underestimated. Increasing the mass loss rates shortens the period preceding the time when layers having belonged to the convective core appear at the surface and increases the quantity of mass lost enriched in $^{26}$Al from that stage on. 
From Fig.~\ref{fig:kipp}, looking at the 250 M$_\odot$ model, we see that
the first effect, \textit{i.e.} shortening the phase before the surface is $^{26}$Al enriched, will have little impact. Indeed, this phase is already very short.
With the present mass loss rates, starting from the time when the surface is enriched in $^{26}$Al, more than 100 M$_\odot$ are lost through winds. 
{ Increasing the amount of mass lost will increase the $^{26}$Al yield, and will thus increase the contribution of VMS on the global $^{26}$Al budget}.


The impact of rotational mixing is also very significant, as a population of rotating massive stars produces twice the { amount of} $^{26}$Al than a { population of non-rotating stars} would in the Milky Way. An increase of the rotation rate would result in even higher $^{26}$Al yields due to an even stronger transport from rotational mixing, bringing even more $^{26}$Al to the surface early-on.
Of course, less efficient transport of the chemical species will act the other way round and would decrease the quantities of $^{26}$Al ejected by stellar winds.
Convection also plays an important role in the transport of $^{26}$Al to the surface. With an increase of the core size \citep[\textit{e.g} as suggested by][]{Martinet2021,Scott2021}, the transport of $^{26}$Al would also be enhanced (as we have seen Fig. \ref{fig:kipp}), resulting in larger $^{26}$Al yields.
We have seen that increasing the initial mass increases the mass loss and the convective core size for VMS. It means that increasing the upper mass limit for VMS (\textit{e.g} 300 to 500 \Msol) would also result in an even larger increase of the $^{26}$Al galactic production. This can be seen in Fig. \ref{fig:upper_mass_limit}, where the $^{26}$Al production in the galaxy is displayed for an IMF with an upper-mass limit up to 500 \Msol. The yields used { for stars with initial masses} over 300 \Msol\ have been extrapolated and are compatible with the mass loss obtained in the 500 \Msol\ models of \citet{Yusof2013}. Pushing the upper-mass limit to 500 \Msol\ increases the total $^{26}$Al output from winds by { a factor of} two in comparison to 300 \Msol\ and would account for a very large fraction of the galactic $^{26}$Al.

Some changes in the nuclear reaction rates may of course affect the present results. The interested reader can look at the paper by \citet{Ilia2011}, in which a detailed study of the dependence of the $^{26}$Al masses synthesized by massive stars on the nuclear reaction rates has been performed. For the $^{26}$Al produced in the H-burning convective core, an important reaction is the one synthesizing 
$^{26}$Al from $^{25}$Mg { by the $^{25}$Mg(p,$\gamma$)$^{26}$Al reaction}. A higher reaction rates will
increase the quantity of $^{26}$Al ejected by the winds, and the contrary for a lower rate. The median rate by \citet{Illiadis2010} for the $^{25}$Mg(p,$\gamma$)$^{26}$Al 
is at most 20\% smaller than the rate used in this work
\citep[taken from][]{Iliadis2001} for the typical temperatures in the H-burning cores of massive stars. Such a reduction
will not have a dramatic impact on the yields, lowering them slightly if { all other parameters are unchanged}. Very recently, the $^{26}$Al(n,$\alpha$) reaction rate has been updated by \citet{Ledere2021}, but this reaction does not impact the phase during which most of the $^{26}$Al is produced and ejected and thus it should not have a strong effect on the present results. 

\subsection{Synthesis of the main results and future perspectives}
\label{sect:conlusion_perspectives}
\begin{figure}
    \centering
     \includegraphics[width=0.49\textwidth]{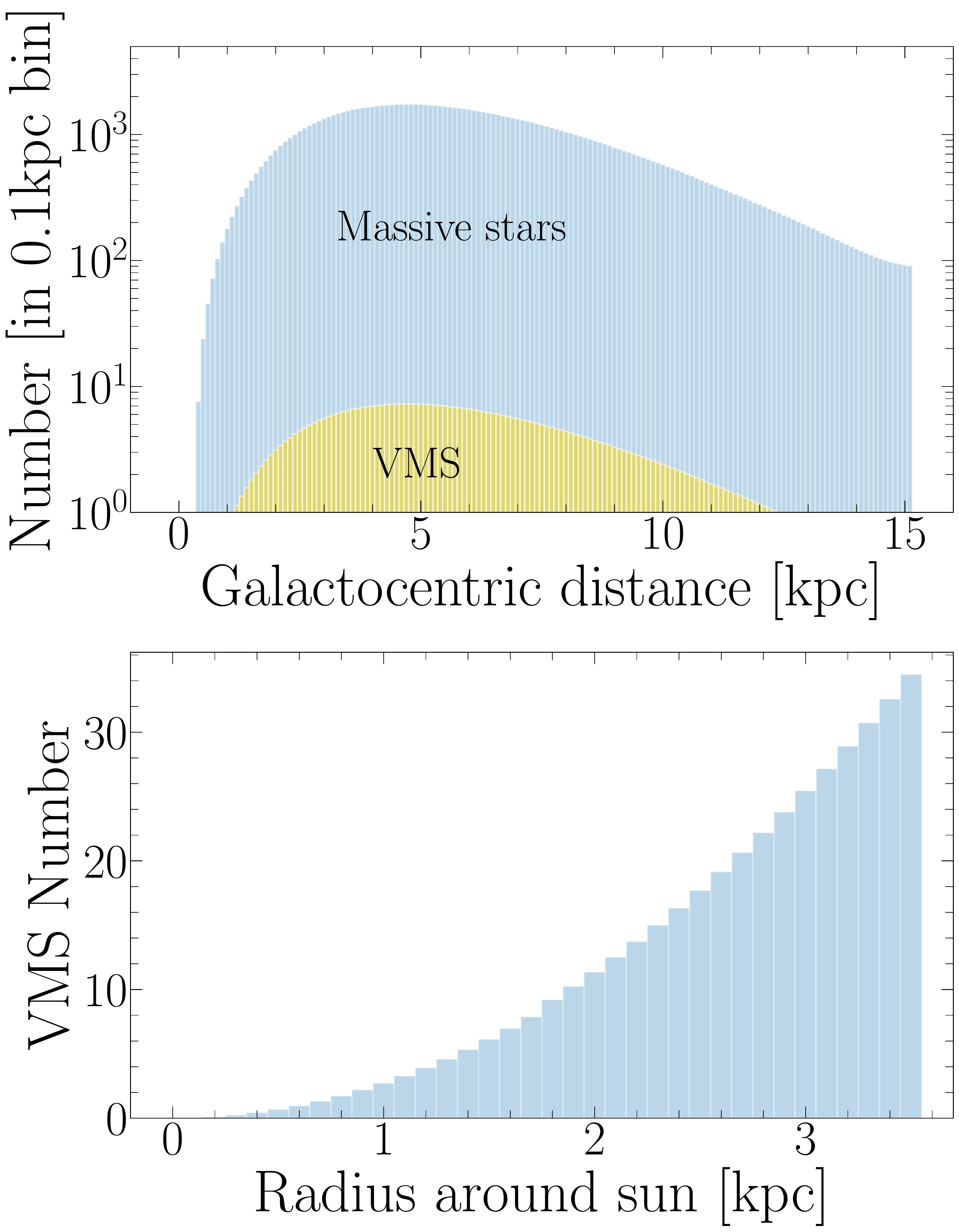}
     \caption{\textit{Top panel:} Number of massive stars (8-120 \Msol) and VMS (120-300 \Msol) in rings of 0.1 kpc width as a function of the galactocentric distance. We use here a Salpeter's IMF with $\alpha$=2.35 and the SFR used in the middle panel of Fig. \ref{fig:Al26_galaxy} from \citet{Kubryk2015}. \textit{Bottom panel:} Number of VMS in a disk around the sun as a function of the maximal distance to the sun, following the same methodology.  }
     \label{fig:Number_VMS_galaxy}
\end{figure}

We computed the { yields} of $^{26}$Al ejected by stellar winds for massive and very massive stars with and without rotation at three different metallicities. We showed that their impact on the global budget of $^{26}$Al is significant.
This underlines the need to search for such objects and obtain data on their frequency at different metallicities/environments. 

Using a simple galactic model to compute the total $^{26}$Al mass due to the winds of massive and very massive stars, we obtain that the Galaxy globally contains around 120 000 stars with masses between 8 and 300M$_\odot$, and 500 with initial masses between 120 and 300M$_\odot$.
The variation of these numbers as a function of the galactocentric distance
is shown in Fig.~\ref{fig:Number_VMS_galaxy} (top panel). The lower panel displays the expected number of VMS in a disk around the sun as a function of the radius of the disk. For example, we expect about 2.5 VMS in a sphere of 1 kpc radius around the Sun. These are of course very rough estimates.
Figure~1 in \citet{Meynet1994}
gives the maximum distance at which a given mass of $^{26}$Al can be detected as a point source by INTEGRAL. If one considers the mass ejected by our rotating model of 180 \Msol\ and assume that half of the total yield of 1.3e-03 M$_\odot$ is still emitting { $\gamma$-ray emission at 1.8 MeV} and that this amount
is sufficiently concentrated that it would appear as a point source, then
the maximum distance at which such a star would be detectable is 1.25 kpc.
This is a very optimistic value. { Indeed, stochastic effects, the fact that such a source might not appear as point-like or (that the source) presents a smaller amount of
non-decayed $^{26}$Al than assumed here, would decrease the distance at which the star could be detectable.}
At the moment, the most convincing candidate for very massive stars 
have been observed in the LMC \citep{Crowther2010,Bestenlehner2020}. They are way too far (roughly 50 kpc) for being observed at 1.8 MeV. 
An interesting candidate in the Galaxy is  Westerhout 49-2 \citep{Wu2016}, that has an estimated mass of 250\Msol, but lies at 11.1 kpc, still far beyond INTEGRAL sensitivity range. One { of the closest} candidates would be WR 93 (HD 157504), a 120 \Msol\ star \citep{Rate2020}. Its distance of 1.76 kpc is however beyond the maximum distance at which such a source might be detectable.
This discussion indicates that while VMS might contribute significantly to the enrichment, it is still compatible with the fact that no point-like source of $^{26}$Al has been detected so far. 
While the detection of such a source seems plausible in terms of instrument sensitivity, it would necessitate highly favorable circumstances that seem, for the time being, unlikely.

{ In addition to the question of the origin of the galactic $^{26}$Al,
this isotope is also much discussed in two other contexts.
There is ample evidence in meteorites for the presence of live $^{26}$Al { in the cloud that gave birth to the solar system 4.56 Gyr ago} \citep[\textit{e.g.}][]{Lee1976,Park2017}. Because of its short half-life, $^{26}$Al can help us probe the Solar System astrophysical environment at its birth \citep[\textit{e.g.}][]{Adams2001}. To account for $^{26}$Al elevated abundance in the nascent Solar System compared to the $^{26}$Al/$^{27}$Al ratio found in Calcium-aluminium-rich inclusions \citep{Jacobsen2008}, in-situ irradiation of the protoplanetary disk gas and dust has been proposed \citep{Lee1977,Gounelle2001} but is now abandoned after some nuclear cross-sections have been remeasured \citep{Fitoussi2008}. Late delivery by supernovae and AGB stars has long been considered as a possibility to account for solar System $^{26}$Al. Supernovae are now often discarded because they would yield a $^{60}$Fe/$^{26}$Al ratio at least one order of magnitude larger than the one observed in the early solar system \citep{Gounelle2008}. The probability of associating an AGB star with a star-forming region is very small \citep{Kastner1994}. At present, the most promising models are the ones involving { the winds of} massive stars \citep{Arnould1997,Gounelle2012}. A setting whereby a massive star injects $^{26}$Al in a dense shell it itself generated \citep{Gounelle2012,Dwar2017} seems to be a likely model \citep{Gounelle2015}.


Another aspect in which $^{26}$Al plays an important role is the question of the sources of pristine circumstellar grains, \textit{i.e.} of grains
formed around stars that travel through space and are
finally locked in meteorites.
Traces of $^{26}$Mg due to $^{26}$Al present at the time of formation of these pristine circumstellar grains, along with the measurements of the abundances of other isotopes, provide a clue about the nature of stars that produce such grains. \citep[see e.g.][]{Zinner1991,Dauphas2011}.
What proportion of these grains might be produced around VMS, or might the
$^{26}$Al injected into the proto-solar nebula comes from a very massive star, are interesting questions for future prospects.}

\begin{acknowledgements}
SM has received support from the SNS project number 200020-205154.
GM, SE, CG and DN have received funding from the European Research Council (ERC) under the European Union's Horizon 2020 research and innovation program (grant agreement No 833925, project STAREX).
RH acknowledges support from the World Premier International Research Centre Initiative (WPI Initiative, MEXT, Japan), STFC UK, the European Union’s Horizon 2020 research and innovation program under grant agreement No 101008324 (ChETEC-INFRA) and the IReNA AccelNet Network of Networks, supported by the National Science Foundation under Grant No. OISE-1927130. NY acknowledged the support from Fundamental Research Grant scheme grant number FP042-2021 under Ministry of Higher Education Malaysia. 
VVD's work is supported by National Science Foundation award 1911061 to
the University of Chicago (PI V. Dwarkadas)
\end{acknowledgements}

\bibliographystyle{aa} 
\bibliography{GRID006} 

\appendix

\newpage
\section{Table}

Table \ref{tab:Al26_production} presents the stellar $^{26}$Al yields from winds obtained in this work as a function of the initial mass and metallicity, for both non-rotating and rotating at V/V$_c$=0.4 models.

\begin{table*}
    \centering
    \caption{$^{26}$Al wind yields (calculated using Eq. \ref{Eq:stellar_yields}) in \Msol\ units. { The VMS models at Z=0.006 and Z=0.014 are from Martinet et al. (in prep.) probing initial masses of 180, 250 and 300\Msol, while the models at Z=0.020 are from \citet{Yusof2022}, with VMS models computed only for 150 and 300\Msol. { More details on the massive stars models at Z=0.006 can be found in \citet{Eggenberger2021} and in \citet{Ekstrom2012} for the ones at Z=0.014}. Dashes indicate where the models where not computed.} }
\begin{tabular}{r|rr|rr|rr}
\toprule
\toprule
&\multicolumn{2}{c}{Z=0.006}&\multicolumn{2}{c}{Z=0.014}&\multicolumn{2}{c}{Z=0.020}\\
 Mass & Non-rotating &  V/V$_c$=0.4 &  Non-rotating &  V/V$_c$=0.4 &  Non-rotating &  V/V$_c$=0.4 \\
\midrule
   12 \Msol &          4.63E-13 &      3.85E-11 &          1.01E-09 &      4.62E-09 &          2.09E-09 &      1.46E-08 \\
   15 \Msol &          1.94E-12 &      7.73E-10 &          1.48E-10 &      3.09E-08 &          1.12E-09 &      1.00E-07 \\
   20 \Msol &          1.12E-09 &      4.87E-07 &          5.57E-07 &      4.46E-06 &          1.32E-06 &      5.42E-06 \\
   25 \Msol &          1.89E-06 &      1.06E-05 &          9.50E-06 &      9.67E-06 &          1.09E-05 &      2.02E-05 \\
   32 \Msol &          1.67E-05 &      1.24E-05 &          2.04E-05 &      3.08E-05 &          2.72E-05 &      4.25E-05 \\
   40 \Msol &          2.32E-05 &      1.53E-05 &          3.82E-05 &      3.66E-05 &          4.03E-05 &      6.73E-05 \\
   60 \Msol &          7.70E-05 &      2.24E-05 &          5.33E-05 &      1.05E-04 &          1.12E-04 &      2.27E-04 \\
   85 \Msol &          4.25E-05 &      5.17E-05 &          1.67E-04 &      3.21E-04 &          2.78E-04 &      5.68E-04 \\
  120 \Msol &          9.57E-05 &      1.39E-04 &          3.64E-04 &      7.09E-04 &          6.50E-04 &      1.26E-03 \\
  150 \Msol &               - &           - &               - &           - &          1.17E-03 &      1.96E-03 \\
  180 \Msol &          1.70E-04 &      3.37E-04 &          8.37E-04 &      1.30E-03 &               - &           - \\
  250 \Msol &          3.67E-04 &      6.51E-04 &          1.93E-03 &      2.63E-03 &               - &           - \\
  300 \Msol &          5.39E-04 &      8.97E-04 &          3.03E-03 &      3.71E-03 &          5.78E-03 &      6.43E-03 \\
\bottomrule
\end{tabular}

    \label{tab:Al26_production}
\end{table*}

\end{document}